# A New Quasi One-Dimensional Compound Ba$_3$TiTe$_5$ and Superconductivity Induced by Pressure

Jun Zhang[1,2,3†], Yating Jia[1,2†], Xiancheng Wang*[1], Zhi Li[4], Lei Duan[1,2], Wenmin Li[1,2], Jianfa Zhao[1,2], Lipeng Cao[1], Guangyang Dai[1,2], Zheng Deng[1], Sijia Zhang[1], Shaomin Feng[1], Ruize Yu[1], Qingqing Liu[1], Jiangping Hu[1,2], Jinlong Zhu*[5,3] and Changqing Jin*[1,2,6]

[1]*Beijing National Laboratory for Condensed Matter Physics, Institute of Physics, Chinese Academy of Sciences, Beijing 100190, China.*

[2]*School of Physics, University of Chinese Academy of Sciences, Beijing 100190, China*

[3]*Center for High Pressure Science & Technology Advanced Research, Beijing 100094, China.*

[4] *College of Materials Science and Engineering, Nanjing University of Science and Technology, Nanjing 210094, China*

[5] *Physics Department, Southern University of Science and Technology, Shenzhen 518055, China*

[6]*Materials Research Lab at Songshan Lake, 523808 Dongguan, China*

*Corresponding author: wangxiancheng@iphy.ac.cn; jinlong.zhu@hpstar.ac.cn; jin@iphy.ac.cn

† These authors contributed equally to this work

Keywords: Quasi one-dimension; Conducting chains; High pressure; Superconductivity

## Abstract

We report systematical studies of a new quasi-one-dimensional (1D) compound Ba$_3$TiTe$_5$ and the high-pressure induced superconductivity therein. Ba$_3$TiTe$_5$ was synthesized at high pressure and high temperature. It crystallizes into a hexagonal structure ($P6_3/mcm$), which consists of infinite face-sharing octahedral TiTe$_6$ chains and Te chains along the $c$ axis, exhibiting a strong 1D characteristic structure. The first-principles calculations demonstrate that Ba$_3$TiTe$_5$ is a well-defined 1D conductor and thus, it can be considered a starting point to explore the exotic physics induced by pressure via enhancing the interchain hopping to move the 1D conductor to a high dimensional metal. For Ba$_3$TiTe$_5$, high-pressure techniques were employed to study the emerging physics dependent on interchain hopping, such as the Umklapp scattering effect, spin/charge density wave (SDW/CDW), superconductivity and non-Fermi Liquid behavior. Finally, a complete phase diagram was plotted. The superconductivity emerges from 8.8 GPa, near which the Umklapp gap is mostly suppressed. $T_c$ is enhanced and reaches the maximum ~6 K at about 36.7 GPa, where the spin/charge density wave (SDW/CDW) is completely suppressed, and a non-Fermi Liquid behavior appears. Our results suggest that the appearance of superconductivity is associated with the fluctuation due to the suppression of Umklapp gap and the enhancement of $T_c$ is related with the fluctuation of the SDW/CDW.

## Introduction

The 1D system has attracted much attention due to its novel physics and unique phenomena, which are dramatically different from two- or three-dimensional systems[1, 2]. When the motion of electrons is confined within 1D, the electrons cannot move without pushing all the others, which leads to a collective motion and thus, spin-charge separation. In such a case, the concept of "quasi-particles" with charge and spin degrees of freedoms is replaced with the collective modes, and the electronic state in 1D system is predicted by Tomonaga-Luttinger liquid (TLL) theory. Pressure is a unique technique to tune the interchain hopping to gradually transform the 1D conductor to high dimensional metal (HDM), during which many interesting physical phenomena emerge, such as superconductivity. The properties of quasi-1D conductors dependent on the strength of interchain coupling have been extensively explored in the organic compounds, such as $(TMTTF)_2X$ and $(TMTSF)_2X$ salts, which exhibit a 1D conducting characteristic with the overlap integrals ratio of different axis $t_a:t_b:t_c \sim 200:20:1$ [3-7]. However, the pressure-temperature phase diagrams were only plotted in a narrow range of pressure (generally less than 2 GPa) for each quasi-1D organic compound. For $(TMTTF)_2X$ salts, the conducting chains along the $a$ axis are less coupled. When temperature decreases, they successively undergo metal-insulator transition induced by Umklapp scattering (U-MIT) and ordered phase transition such as charge order, both of which can be gradually suppressed by pressure. While for $(TMTSF)_2X$ salts, the enhancement of interchain coupling $t_b$ completely suppresses the U-MIT. Within the perspective, $(TMTSF)_2X$ can be considered as the high pressure phase of $(TMTTF)_2X$. At low temperature, $(TMTSF)_2X$ salts exhibit HDM behavior due to the single-particle interchain hopping. For $(TMTSF)_2PF_6$, a spin density wave (SDW) state forms in the HDM region [8]. With applying pressure, the SDW transition is suppressed gradually and superconductivity is induced [9-11]. When the interchain coupling further increases, such as in $(TMTSF)_2ClO_4$, superconductivity appears with the complete suppression of the SDW [12].

Besides the organic system, the inorganic quasi 1D conductors have also received considerable attention. For example, in the compound $Li_{0.9}Mo_6O_{17}$ the ratio of conductivity along the $b$-, $c$-, and $a$-axes is about 250: 10: 1 [13]. It has been evidenced to be a TLL state in high-temperature region[14-16]. $Li_{0.9}Mo_6O_{17}$ undergoes a dimensional crossover from a 1D conductor to 3D metal at ~24 K. The dimensional crossover destabilizes the TLL fixed point, induces an electronic SDW/CDW and thus, leads to a crossover from metal to semiconductor[17]. By decreasing the temperature further, $Li_{0.9}Mo_6O_{17}$ exhibits a superconducting transition at 1.9 K[13, 17-20]. $M_2Mo_6Se_6$ (M=Rb, Na, In and Tl) is another interesting quasi-1D system with $4d$ transition metal, which consists of conducting $(Mo_6Se_6)$ chains along the $c$ axis, and the chains are separated by M cations in the $ab$-plane [21-24]. The interchain coupling is controlled by the size of M cations and increases with the sequence of Rb, Na, In and Tl. $Rb_2Mo_6Se_6$ undergoes a CDW transition at about 170 K; while for $Na_{2-\delta}Mo_6Se_6$, $In_2Mo_6Se_6$, and $Tl_2Mo_6Se_6$, superconductivity appears with $T_c$ about 1.5 K, 2.8 K, and 4.2 K, respectively[21, 24]. Recently, the remarkable quasi 1D superconductor of $K_2Cr_3As_3$ with $T_c$~6.1 K and its related superconductors of Cr-233-type $(Na/Rb)_2Cr_3As_3$ and Cr-133-type $(Rb/K)Cr_3As_3$ have been reported[25-29]. Their conducting chains are double-walled subnanotubes $[(Cr_3As_3)^{2-}]_\infty$ along the $c$ axis and separated by alkali metal. It is interesting that the superconducting $T_c$ decreases monotonously with the increase of the distance between the adjacent conducting chains of $[(Cr_3As_3)^{2-}]_\infty$ in Cr-233-type materials, which is tuned by the radius of the cations of alkali metal [27]. Anyway, for inorganic quasi 1D conductor, the superconductivity dependent on interchain coupling has been less studied.

Another important series is $R_3TiSb_5$ ($R$ = La, Ce) due to its strong 1D structure characteristics, which consist of face-sharing octahedral $TiSb_6$ chains and Sb-chains, and these chains are separated by $R$ atoms [30-32]. If ignoring the contribution of the La atoms, the band structure calculation on the $[TiSb_5]^{9-}$ substructure of $La_3TiSb_5$ compound suggests a well-defined 1D conductor [30]. However, the complementary calculation proves that there are non-negligible contributions of La to the density of state (DOS) at the Fermi level. The $La^{3+}$ ions in $La_3TiSb_5$ are not perfectly ionic, which leads to a 3D band structure [32]. It is strongly indicated that La atoms do not play the role of interchain separation but serve as a bridge for electrons hopping among the chains.

It is important to find an ideal 1D conductor with a simple structure to systematically explore the interchain hopping modulated Umklapp gap, the emerging SDW/CDW and superconductivity. Here, we used the metastable compound $Ba_3TiTe_5$ via the substitution of the rare-earth metal La in $La_3TiSb_5$ by alkali earth metal Ba, and for the charging compensation, the group $V_A$ element of Sb was replaced with the group $VI_A$ element of Te, which was synthesized under high-pressure and high-temperature (HPHT) conditions. The structure of $Ba_3TiTe_5$ consists of infinite octahedral $TiTe_6$ chains and Te chains along the $c$ axis, and our calculation proved its band structure has a well-defined 1D conducting characterization. On this basis, to explore the sequence emergent phenomena, we further employed high pressure, which is clean without impurity disturbance and can also effectively and continuously tune the interchain hopping. Finally, we plotted a complete phase diagram with the single material of $Ba_3TiTe_5$, and within a wide pressure range to present the evolution from 1D conductor to HDM and the emergent physics, during which the pressure-induced superconductivity was observed. Our results indicate that the fluctuation due to the suppression of Umklapp gap and SDW/CDW responds to the appearance of superconductivity and the enhancement of $T_c$, respectively.

## Materials and Method

### Materials and Synthesis
The lumps of Ba (Alfa, immersed in oil, >99.2% pure), Te powder (Alfa, >99.99% pure) and Ti powder (Alfa, >99.99% pure) were purchased from Alfa Aesar. The precursor BaTe was prepared by reacting the lumps of Ba and Te powder in an evacuated quartz tube at 700℃. $Ba_3TiTe_5$ was synthesized under HPHT conditions. The obtained BaTe powder, Te, and Ti powder were mixed according to the elemental ratio of stoichiometric $Ba_3TiTe_5$, and ground and pressed into a pellet. The pre-pressed pellet was treated using a cubic anvil high-pressure apparatus at 5 GPa and 1300 ℃ for 40 mins. After the high-pressure and high-temperature process, the black polycrystalline sample of $Ba_3TiTe_5$ was obtained.

### Measurements
The ambient X-ray diffraction was conducted on a Rigaku Ultima VI (3KW) diffractometer using Cu $K_\alpha$ radiation generated at 40 kV and 40 mA. The *in situ* high-pressure angle-dispersive X-ray diffraction was collected at the Beijing Synchrotron Radiation Facility at room-temperature with a wavelength of 0.6199 Å. Diamond anvil cell with 300 $\mu$m cullet was used to produce high pressure, and silicone oil was used as the pressure medium. The Rietveld refinements on the diffraction patterns were performed using GSAS software packages[33]. The crystal structure was plotted with the software VESTA[34].

The *dc* magnetic susceptibility measurement was carried out using a superconducting quantum interference device (SQUID). The resistance was measured by four-probe electrical conductivity methods in a diamond anvil cell made of CuBe alloy using a Mag lab system. The diamond culet was 300 $\mu$m in diameter. A plate of T301 stainless steel covered with *c*-BN powder was used as a gasket, and a hole of 150 $\mu$m in diameter was drilled in the pre-indented gasket. Fine *c*-BN powders were pressed into the holes and further drilled to 100 $\mu$m, serving as the sample chamber. Then, NaCl powder was put into the champ as a pressure-transmitting medium, on which the $Ba_3TiTe_5$ sample with a dimension of 60 × 60 × 15 $\mu m^3$ and a tiny ruby were placed. The pressure was calibrated using the ruby florescent method. At each pressure point, the anvil cell was loaded into the Mag lab system for the transporting measurement with an automatically controlled temperature and magnetic field.

**Calculations**

The first-principles calculations based on density functional theory implemented in VASP were carried out within a primitive cell with an 8×8×16 k-point grid[35]. The projector augmented wave pseudopotentials with Perdew, Burke, and Ernzerhof (PBE) exchange-correlation and 450 eV energy cutoff were used in our calculation[36,37]. The experimental lattice parameters obtained from XRD were adopted.

**Results and Discussion**

The X-ray diffraction of $Ba_3TiTe_5$ at ambient condition is shown in Fig. 1(a). All the peaks can be indexed by a hexagonal structure with lattice constants *a* = 10.1529 Å and *c* = 6.7217 Å, respectively. The space group of *P*6$_3$/*mcm* (193) is used according to the systematic absence of *hkl*. Finally, the Rietveld refinement was performed by adopting the crystal structure of $R_3TiSb_5$ (*R* = La, Ce) as the initial model [30,31], which smoothly converged to $\chi^2$ = 2.0, *Rp* = 3.1% and *Rwp* = 4.4%. The summary of the crystallographic data is listed in Table I.

The schematic plot of the crystal structure is shown in Fig. 1(b, c). Fig. 1(b) is the top view with the projection along the *c* axis, displaying the triangular lattice form; while Fig. 1(c) is the side view to show the chain geometry. From Fig. 1(b, c), we can see that the crystal structure of $Ba_3TiTe_5$ consists of infinite face-sharing octahedral $TiTe_6$ chains and Te chains along the *c* axis, which are separated by Ba cations. The distance between the adjacent $TiTe_6$ chains is given by the lattice constant of *a*=10.1529 Å, which is significantly large, and responsible for exhibiting a quasi-1D structure characteristic. To explore the 1D characteristics of $Ba_3TiTe_5$ from a band structure perspective, we calculated the band structure, partial density of state (PDOS) and Fermi surface via first-principles calculations for $Ba_3TiTe_5$ under ambient pressure, as shown in Fig.1(d-f). The hallmark of the band structure is that the bands with the k-point path parallel to the $k_z$-direction intercept the Fermi level; while for k-point path perpendicular to the $k_z$-direction, the band dispersion is always gapped. Thus, $Ba_3TiTe_5$ is a well-defined 1D conductor with the conducting path along the z-direction. Additionally, from the PDOS, it can be seen that the DOS near the Fermi level is dominated by the 3*d*-orbitals of Ti. The 5*p*-orbitals of both Te(1) and Te(2) from the $TiTe_6$ chains and Te chains, respectively, contribute to the DOS near the Fermi level as well, which proposes that both the TiTe$_6$ chains and Te chains are conducting chains. The DOS from Ba 6s-orbital, presented in Fig. 1(e) as well, is close to zero at Fermi level and can be ignored, which suggests that the conducting chains are well separated by Ba$^{2+}$ ions. Fig. 1(f) displays the calculated Fermi surfaces. There are four sheet-like Fermi surfaces perpendicular to the $k_z$-direction, and the bottom sheet can be shifted by the wave vectors $k_1$ and $k_2$ to nest with the above two sheets, respectively. Therefore, the Fermi surfaces are unstable and the transport property of 1D conductor should be described by TLL theory. The resistivity measurement at ambient pressure for Ba$_3$TiTe$_5$ was carried out, as shown in Fig. 1(g). The resistivity increases as the temperature decreases, exhibiting a semiconducting behavior. The inset is the ln($\rho$) versus reverse temperature. By fitting the ln($\rho$)-1/$T$ curve according to the formula of $\rho \propto \exp(\Delta_g/2k_BT)$, where $k_B$ is the Boltzmann's constant, the band gap of $\Delta_g$ can be estimated to be 232 meV. For a 1D conducting system, Umklapp scattering has important influence on the electron transfer, which usually results in a correlation gap and insulating state [38-41]. Beside the Umklapp scattering effect, non-zero disorder in 1D conducting system tends to localize the electrons as well. If the disorder dominates the localization, the electron transport should be described by the model of various range hopping. Here, the resistivity following the Arrhenius law within the measured temperature range proves that the Umklapp scattering effect should play the role for the metal-insulating transition, since the Umklapp process produces a real correlation gap. Therefore, the inconsistency between the measured resistivity and the calculated results should arise from the Umklapp scattering effect. In addition, the magnetic susceptibility measurement shows that Ba$_3$TiTe$_5$ is non-magnetic in the measured temperature range from 2 K to room temperature, as shown in Fig. S1.

High pressure is an effective way to tune the lattice of a crystal structure. In the 1D system, it can significantly decrease the distance between adjacent conducting chains, thus, enhance the interchain hopping and move the 1D conductor to HDM, during which rich interesting physics are induced, such as the SDW/CDW and superconductivity. Therefore, continually compressing a 1D conductor can provide a potential pathway to understanding the rich phase diagram and the fundamental underlying mechanism. Here, high-pressure X-Ray diffraction experiments for Ba$_3$TiTe$_5$ were performed first to study the structural stability and the pressure dependence of the lattice parameters, as shown in Fig. S2-S8. Within the highest measured pressure of 50.6 GPa, the hexagonal structure of Ba$_3$TiTe$_5$ is stable, and the distance between the adjacent conducting chains decreases gradually by 12.1%, which is ideal to explore the exotic emergent physics dependent on the interchain hopping tuned by pressure. Therefore, we carried out the resistance measurements under high pressure, as shown in Fig. 2(a,b). Although the resistance decreases with increasing pressure when the pressure is lower than 7 GPa, it is still very high (~10$^6$ Ω at 6.7 GPa and 2 K). At the pressure of 8.8 GPa, the resistance drops dramatically by four orders of magnitude down to ~10-100 Ω, which suggests that the Umklapp scattering induced gap should be mostly suppressed. A closer view at this pressure shows there is a downward trend in the low-temperature region of the resistance curve (seen in Fig. S9(a, b)). The downward transition temperature increases from ~5 K to ~7.5 K with initial pressure increases, and then decreases to ~5.4 K at 15.9 GPa. When increasing pressure further, the downward transition temperature begins to increase again, and the downward behavior gradually develops into a superconducting transition, which persists to the highest measured pressure of 58.5 GPa. Therefore, we suggest that the superconductivity appearance is associated with the fluctuation due to the suppression of the Umklapp gap. In addition, there is an unknown hump independent of pressure at around 150 K, as shown in Fig. 2(b), which has nothing to do with the superconducting transition and therefore, was not discussed in this work. Fig. 2(c) displays the superconducting transition dependent on the magnetic field at 17.3 GPa. At zero field, the onset transition temperature is about 3.9 K, and the resistance drops to zero at ~2.8 K. Upon applying the magnetic field, the transition is gradually suppressed. Taking the criterion of the onset transition as the superconducting transition temperature, the curve of $H_{c2}$ versus $T_c$ is plotted in the inset of Fig. 2(c), where the slope of $-dH_{c2}/dT|_{Tc}$ is ~3.07 T/K. Using the Werthamer-Helfand-Hohenberg formula $\mu_0H_{c2}^{Orb}(0) = -0.69 \,(dH_{c2}/dT)\, T_c$ and taking $T_c = 3.8$ K [42], the upper critical field limited by the orbital mechanism is estimated to be $\mu_0H_{c2}^{Orb}(0) \sim 8$ T. Another mechanism determining the upper critical field is the Pauli paramagnetic effect. The upper critical field is estimated with the formula $\mu_0H_{c2}^{P}(0) = 1.84\, T_c \approx 7$ T [43], which is comparable with $\mu_0H_{c2}^{Orb}(0)$.

The enlarged view of the resistance curves is plotted in Fig. 3(a,b) and Fig. 3(d,e) to display more detailed information. The resistances below 36.7 GPa are normalized by $R(150K)$. Above 18.9 GPa, besides the superconducting transition, a metallic state starts to develop at relatively high temperature and is then followed by a resistance upturn while temperature decreases, forming a resistance minimum at $T_m$ marked by the arrow shown in Fig. 3(a). $T_m$ shifts to low temperature with increasing pressure. At 28.7 GPa, the metal-semiconductor crossover (MSC) $T_m$ was determined by the minimum value of the temperature derivative of resistance $dR/dT$ (shown in Fig. S10). Although the crossover temperature cannot be unambiguously determined when the pressure exceeds 28.7 GPa, it is clear that the upturn is completely suppressed at 36.7 GPa. Fig. 3(b) shows the superconducting transition below 36.7 GPa. The $T_c$ increases from 4.3 K to 6.4 K when pressure increases from 18.9 GPa to 36.7 GPa. $T_c$ and $T_m$ as a function of pressure are plotted in Fig. 3(c); when pressure increases, $T_m$ decreases while $T_c$ increases. The $T_c$ reaches the maximum at the critical pressure of 36.7 GPa, where the MSC is completely suppressed.

The phenomenon of MSC has been reported in the quasi-1D compounds of $Li_{0.9}Mo_6O_{17}$ and $Na_{2-\delta}Mo_6Se_6$ [15, 17-19, 21, 44]. Several mechanisms can cause the MSC or MIT, such as Mott instability, SDW/CDW formation, and disorder-induced localization. For $Li_{0.9}Mo_6O_{17}$, the MSC can be gradually suppressed and tuned to be metallic by the magnetic field, suggesting the MSC is the consequence of the SDW/CDW gap (less than 1 meV) formation [19]. Above $T_m$, $Li_{0.9}Mo_6O_{17}$ is evidenced to exhibit TLL behavior [15, 16]. A dimensional crossover happens at $T_m$ and causes the destabilization of the TLL fixed point, leading to the formation of an electronic SDW/CDW, which is suggested to be the origin of MSC in $Li_{0.9}Mo_6O_{17}$ [17]. While for $Na_{2-\delta}Mo_6Se_6$, the MSC temperature $T_m$ is sample dependent and ranges from 70 K to 150 K due to a small variation of Na stoichiometry. It is speculated that the MSC arises from the localization induced by disorder [21]. For an ideal 1D conducting system, the Fermi surface is unstable and the system is TLL state. When increasing the interchain hopping to move the 1D system towards HDM, the Fermi surface nesting established in the quasi 1D conducting system can usually give a SDW/CDW transition and open a gap. In the case of $Ba_3TiTe_5$ under high pressure, for example at 19.5 GPa, the Fermi surface nesting can be observed, as will be discussed in the following. Therefore, the MSC found in $Ba_3TiTe_5$ under high pressure is suggested to arise from the SDW/CDW transition, and the MSC $T_m$ in Fig. 3(a, c) should correspond to the SDW/CDW transition temperature. The $T_c$ increases with SDW/CDW suppression and reaches the maximum when the SDW/CDW transition is suppressed to zero. It is speculated that the superconductivity is enhanced by the fluctuation of

SDW/CDW.

Fig. 3(d,e) shows the temperature dependence of resistance with pressure exceeding 36.7 GPa. There is an obvious hump below the onset $T_c$. The resistance curve demonstrates a two-step superconducting transition. In fact, the two-step superconducting transition has been reported in other quasi-1D superconductors, where the onset transition is ascribed to the superconducting fluctuation along individual chains, and the hump signifies the onset of transverse phase coherence due to the interchain coupling [21, 45-47]. Here, the lower temperature transition should be attributed to the transverse phase coherence since the two-step transition feature becomes more pronounced when the interchain coupling is enhanced by pressure. The $T_c$ versus pressure in this pressure region is plotted in Fig. 3(f). The $T_c$ monotonously decreases with further increasing pressure. The normal state of resistance between 10 K and 60 K is fitted by the formula $R=R_0+AT^n$, as shown in Fig. 3(d), where $R_0$ is the residual resistance, $A$ is the coefficient of the power law, and $n$ is the exponent. The $R_0$ value ranges from 0.07 Ω to 0.11 Ω. The size of our sample for high pressure measurements is about 60 μm × 60 μm with the height about 15 μm. Thus, the residual resistivity can be estimated to be 1.0-1.6×10$^{-4}$ Ω-cm, which is comparable with that reported in (TMTTF)$_2$AsF$_6$[7]. The obtained exponent $n$ dependent on pressure is plotted in Fig. 3(f), which shows that $n$ increases from 0.9 to 1.8 as pressure increases, *i.e.*, the system develops from a non-FL to a FL state. To further demonstrate the crossover from a non-FL to a FL state, the temperature and pressure dependent exponent in the metallic region is plotted in Fig. 5, where the color shading represents the value of exponent $n$. The $n$ value approaching 2 near 26.2 GPa should be due to the effect of MSC. When the pressure exceeds 36.7 GPa, the Fermi surface nesting should be broken, as will be discussed in the following, and the FL state develops gradually as pressure increases. It is interesting that the non-FL behavior appears at the critical pressure where the SDW/CDW is wholly suppressed. Therefore, the non-FL behavior may be caused by the SDW/CDW fluctuation. When the system is turned away from the instability of SDW/CDW, the FL state develops gradually. Although the non-Fermi liquid behavior is generally observed in two-dimensional system, it was also reported in organic TMTSF salts, an archetypal quasi-1D system [7, 48, 49]. In fact, the superconductivity of TMTSF salts share the common border with SDW and the magnetic fluctuation gives rise to the linear temperature dependence of resistivity at low temperature [49].

To help understand the above emergent phenomena, we carried out the calculations of band structure, PDOS and Fermi surface for Ba$_3$TiTe$_5$ under different high pressures, which are presented in Fig. 4(a-c) for 19.5 GPa and Fig. 4(d-f) for 42.2 GPa, respectively. For the pressure of 19.5 GPa, the main difference from ambient pressure is that the conduction band bottom around Γ and M points sinks down and just touch the Fermi level and thus, produce small electronic Fermi pockets, which means that the electrons have coherent interchain hopping. These conduction bands are very flat, suggesting the interchain electron mobility is small. Besides the newly formed Fermi pockets, the four Fermi sheets warp slightly, so that the bottom sheet can only roughly nest with the second sheet from the top with the vector $k_2$. According to the electron response function,

$$\chi(q) = \int \frac{dk}{(2\pi)^d} \frac{f_{k+q} - f_k}{E_k - E_{k+q}}$$

where χ(q) is the generalized susceptibility, $f_k$ is the occupation functions of the single-particle states and $E_k$ the single-particle energy, if part of the Fermi surfaces nest, the susceptibility χ(q) should significantly increase and thus, the Fermi surfaces lose the stability, which generally induces a formation of SDW/CDW. Therefore, the MSC observed experimentally in the pressure range of 15-30 GPa should arise from the formation of SDW/CDW induced by Fermi surface nesting. For higher pressure of 42.2 GPa, the band near Γ point further sinks down and crosses the Fermi level, which displays a more 3D-like metal character. The Fermi surfaces become more complex. The Fermi surface around Γ point is more like qusi-1D; while the Fermi surfaces around the A point are between 2D-3D. Thus, the Fermi surfaces loss the nesting and the SDW/CDW are completely suppressed under this pressure, which agrees well with the experimental results. Overall, the ambient 1D electronic state is gradually changed to high dimension but still with anisotropic band structure as the pressure increasing.

The above emergent phenomena induced by pressure are intrinsic to $Ba_3TiTe_5$. First, the possibility of the superconductivity coming from other impurities can be ruled out. Within the X-ray resolution limit, no discernable impurity phase was found in the specimen even in the X-ray diffraction pattern re-plotted with the intensity in logarithmic scale, as shown in Fig. S11. What's more, if there is any impurity containing Ba, Ti, or Te, only the Te is superconducting under high pressure and the pressure dependence of $T_c$ for Te is totally different from $Ba_3TiTe_5$ [50, 51]. Second, the pressure dependence of superconductivity, MSC associated with SDW/CDW and non-Fermi liquid behavior can be reproduced, as shown in Fig. S12 (a-b), which confirms the intrinsic properties of $Ba_3TiTe_5$. Based on the above experiments and discussions, the final temperature-pressure phase diagram of $Ba_3TiTe_5$ is plotted in Fig. 5. At ambient pressure, the quasi-1D conductor $Ba_3TiTe_5$ exhibits semiconducting behavior with a gap of about 232 meV due to the U-MIT. After the suppression of U-MIT, SDW/CDW emerges due to the Fermi surface nesting, which leads to a MSC. Subsequently, the SDW/CDW is gradually suppressed by pressure. Superconductivity appears at 8.8 GPa, where the Umklapp gap has been suppressed completely, and the $T_c$ increases with the suppression of the SDW/CDW. It reaches the maximum of ~6 K at 36.7 GPa, where the normal state of resistance presents a non-FL behavior due to the SDW/CDW fluctuation. With further increasing pressure, the system develops from a non-FL to a FL state since it is away from the SDW/CDW instability. Our results suggest that the pressure-induced superconductivity in quasi-1D conductor $Ba_3TiTe_5$ is initiated by the fluctuation due to the suppression of the Umklapp gap and enhanced by the fluctuation of the SDW/CDW.

## Conclusions

The novel quasi 1D $Ba_3TiTe_5$ conductor was synthesized and extensively studied at high pressure. The conducting paths are the $TiTe_6$ and Te chains, which are separated by Ba cations and thus, present a quasi 1D conducting characteristic. For $Ba_3TiTe_5$, a complete temperature-pressure phase diagram was obtained within a wide pressure range, which presents the evolution from 1D conductor to HDM and the emergent physics. During the increase of pressure, the increased interchain coupling transformed the ambient 1D conductor to high-dimensional metal, during which the pressure induced SDW/CDW, superconductivity, and non-FL behavior appeared. The superconducting transition temperature $T_c$ reaches the maximum accompanied by non-FL behavior when the SDW/CDW gap is suppressed to zero. The superconductivity emergence is closely associated with the suppression of the Umklapp gap and is enhanced by the fluctuation of the SDW/CDW.

## References

1. Voit, J., "One-dimensional fermi liquids", Rep. Prog. Phys. 58, 977 (1995).

2. Giamarchi, T., "Quantum physics in one dimension ", Oxford University Press, Oxford, England (2004).

3. Jerome, D., "Organic conductors: From charge density wave TTF-TCNQ to superconducting $(TMTSF)_2PF_6$", Chem. Rev. 104, 5565 (2004).

4. Kohler, B., *et al.*, "Comprehensive transport study of anisotropy and ordering phenomena in quasi-one-dimensional $(TMTTF)_2X$ salts (X = $PF_6$, $AsF_6$, $SbF_6$, $BF_4$, $ClO_4$, $ReO_4$)", Phys. Rev. B 84, 035124 (2011).

5. Rose, E., *et al.*, "Pressure-dependent structural and electronic properties of quasi-one-dimensional $(TMTTF)_2PF_6$", J. Phys.: Condens. Matter 25, 014006 (2013).

6. Moser, J., *et al.*, "Transverse transport in $(TM)_2X$ organic conductors: possible evidence for a Luttinger liquid", Eur. Phys. J. B 1, 39 (1998).

7. Itoi, M., *et al.*, "Pressure-induced superconductivity in the quasi-one-dimensional organic conductor $(TMTTF)_2AsF_6$", J. Phys. Soc. Jpn. 76, 053703 (2007).

8. Degiorgi, L., *et al.*, "Direct observation of the spin-density-wave gap in (TMTSF)(2)PF6", Phys. Rev. Lett. 76, 3838 (1996).

9. Narayanan, A., *et al.*, "Coexistence of spin density waves and superconductivity in $(TMTSF)_2PF_6$", Phys. Rev. Lett. 112, 146402 (2014).

10. Vuletic, T., *et al.*, "Coexistence of superconductivity and spin density wave orderings in the organic superconductor $(TMTSF)_2PF_6$", Eur. Phys. J. B 25, 319 (2002).

11. Machida, K. & Matsubara, T., "Possibility of coexistence of spin-density wave and superconductivity in organic conductor $(TMTSF)_2PF_6$", Mol. Cryst. Liq. Cryst. 79, 289 (1982).

12. Bechgaard, K., *et al.*, "Superconductivity in an organic-solid - synthesis, structure, and conductivity of Bis(Tetramethyltetraselenafulvalenium) Perchlorate, $(TMTSF)_2ClO_4$", J. Am. Chem. Soc. 103, 2440 (1981).

13. Greenblatt, M., *et al.*, "Quasi two-dimensional electronic-properties of the lithium molybdenum bronze, $Li_{0.9}Mo_6O_{17}$", Solid State Commun. 51, 671 (1984).

14. Denlinger, J. D., *et al.*, "Non-Fermi-liquid single particle line shape of the quasi-one-dimensional non-CDW metal $Li_{0.9}Mo_6O_{17}$: Comparison to the Luttinger liquid", Phys. Rev. Lett. 82, 2540 (1999).

15. Hager, J., *et al.*, "Non-fermi-liquid behavior in quasi-one-dimensional $Li_{0.9}Mo_6O_{17}$", Phys. Rev. Lett. 95, 186402 (2005).

16. Wang, F., *et al.*, "New Luttinger-liquid physics from photoemission on $Li_{0.9}Mo_6O_{17}$", Phys. Rev. Lett. 96, 196403 (2006).

17. Dos Santos, C. A. M., *et al.*, "Dimensional crossover in the purple bronze $Li_{0.9}Mo_6O_{17}$", Phys. Rev. Lett. 98, 266405 (2007).

18. Filippini, C. E., *et al.*, "Pressure effect on the transport-properties of superconducting $Li_{0.9}Mo_6O_{17}$ bronze", Physica C 162, 427 (1989).

19. Xu, X. F., *et al.*, "Directional field-induced metallization of quasi-one-dimensional $Li_{0.9}Mo_6O_{17}$", Phys. Rev. Lett. 102, 206602 (2009).

20. Mercure, J. F., *et al.*, "Upper critical magnetic field far above the paramagnetic pair-breaking limit of superconducting one-dimensional $Li_{0.9}Mo_6O_{17}$ single crystals", Phys. Rev. Lett. 108, 187003 (2012).

21. Petrovic, A. P., *et al.*, "A disorder-enhanced quasi-one-dimensional superconductor", Nat. Commun. 7, 12262 (2016).

22. Potel, M., Chevrel, R. & Sergent, M., "New pseudo-one-dimensional metals - $M_2Mo_6Se_6$ (M = Na, in, K, Tl), $M_2Mo_6S_6$ (M = K, Rb, Cs), $M_2Mo_6Te_6$ (M = in, Tl)", J. Solid State Chem. 35, 286 (1980).

23. Tarascon, J. M., Disalvo, F. J. & Waszczak, J. V., "Physical properties of several $M_2Mo_6X_6$ compounds (M =

Group Ia Metal - X=Se,Te)", Solid State Commun. 52, 227 (1984).

24. Petrovic, A. P., *et al.*, "Phonon mode spectroscopy, electron-phonon coupling, and the metal-insulator transition in quasi-one-dimensional $M_2Mo_6Se_6$", Phys. Rev. B 82, 235128 (2010).

25. Bao, J. K., *et al.*, "Superconductivity in quasi-one-dimensional $K_2Cr_3As_3$ with significant electron correlations", Phys. Rev. X 5, 011013 (2015).

26. Tang, Z. T., *et al.*, "Unconventional superconductivity in quasi-one-dimensional $Rb_2Cr_3As_3$", Phys. Rev. B 91, 020506 (2015).

27. Mu, Q. G., *et al.*, "Ion-exchange synthesis and superconductivity at 8.6 K of $Na_2Cr_3As_3$ with quasi-one-dimensional crystal structure", Phys. Rev. Mater. 2, 034803 (2018).

28. Liu, T., *et al.*, "Superconductivity at 7.3 K in the 133-type Cr-based $RbCr_3As_3$ single crystals", Epl 120, 27006 (2017).

29. Mu, Q. G., *et al.*, "Superconductivity at 5 K in quasi-one-dimensional Cr-based $KCr_3As_3$ single crystals", Phys. Rev. B 96, 140504 (2017).

30. Moore, S. H. D., Deakin, L., Ferguson, M. J. & Mar, A., "Physical properties and bonding in $RE_3TiSb_5$ (RE = La, Ce, Pr, Nd, Sm)", Chem. Mater. 14, 4867 (2002).

31. Bollore, G., Ferguson, M. J., Hushagen, R. W. & Mar, A., "New ternary rare-earth transition-metal antimonides $RE_3MSb_5$ (RE=La,Ce,Pr,Nd,Sm; M=Ti,Zr,Hf,Nb)", Chem. Mater. 7, 2229 (1995).

32. Murakami, T., *et al.*, "Hypervalent bismuthides $La_3MBi_5$ (M = Ti, Zr, Hf) and related antimonides: absence of superconductivity", Inorg. Chem. 56, 5041 (2017).

33. Larson, A. C. & Dreele, R. B. V., "General structure analysis system (GSAS)", Los Alamos National Laboratory Report LAUR 86, 748 (1994).

34. Momma, K. & Izumi, F., "VESTA 3 for three-dimensional visualization of crystal, volumetric and morphology data", J. Appl. Crystallogr. 44, 1272 (2011).

35. Kresse, G. & Hafner, J., "Abinitio molecular dynamics for liquid-metals", Phys. Rev. B 47, 558 (1993).

36. Kresse, G. & Joubert, D., "From ultrasoft pseudopotentials to the projector augmented-wave method", Phys. Rev. B 59, 1758 (1999).

37. Perdew, J. P., Burke, K. & Ernzerhof, M., "Generalized gradient approximation made simple", Phys. Rev. Lett. 77, 3865 (1996).

38. Giamarchi, T., "Umklapp process and resistivity in one-dimensional fermion systems", Phys. Rev. B 44, 2905 (1991).

39. Vescoli, V., *et al.*, "Dimensionality-driven insulator-to-metal transition in the Bechgaard salts", Science 281, 1181 (1998).

40. Pashkin, A., Dressel, M. & Kuntscher, C. A., "Pressure-induced deconfinement of the charge transport in the quasi-one-dimensional Mott insulator $(TMTTF)_2AsF_6$", Phys. Rev. B 74, 165118 (2006).

41. Raczkowski, M. & Assaad, F. F., "Dimensional-crossover-driven mott transition in the frustrated hubbard model", Phys. Rev. Lett. 109, 126404 (2012).

42. Werthamer, N. R., Helfand, K. & Hqhenberg, P. C., "Temperature and purity dependence of the superconducting critical field, $Hc_2$. III. electron spin and spin-orbit effects", Phys. Rev. 147, 295 (1966).

43. Clogston, A. M., "Upper limit for critical field in hard superconductors", Phys. Rev. Lett. 9, 266 (1962).

44. Wu, G. Q., *et al.*, "Direct observation of charge state in the quasi-one-dimensional conductor $Li_{0.9}Mo_6O_{17}$", Sci. Rep. 6, 20721 (2016).

45. Bergk, B., *et al.*, "Superconducting transitions of intrinsic arrays of weakly coupled one-dimensional superconducting chains: the case of the extreme quasi-1D superconductor $Tl_2Mo_6Se_6$", New J. Phys. 13, 103018 (2011).

46. He, M. Q., *et al.*, "1D to 3D dimensional crossover in the superconducting transition of the quasi-one-dimensional carbide superconductor $Sc_3CoC_4$", J. Phys.: Condens. Matter 27, 075702 (2015).

47. Ansermet, D., *et al.*, "Reentrant phase coherence in superconducting nanowire composites", Acs Nano 10, 515 (2016).

48. Doiron-Leyraud, N., *et al.*, "Correlation between linear resistivity and Tc in the Bechgaard salts and the pnictide superconductor $Ba(Fe_{1-x}Co_x)_2As_2$", Phys. Rev. B 80, 214531 (2009).

49. Jerome, D. & Yonezawa, S., "Novel superconducting phenomena in quasi-one-dimensional Bechgaard salts", Comptes Rendus Physique 17, 357 (2016).

50. Akahama, Y., Kobayashi, M. & Kawamura, H., "Pressure-induced superconductivity and phase-transition in selenium and tellurium", Solid State Commun. 84, 803 (1992).

51. Gregoryanz, E., *et al.*, "Superconductivity in the chalcogens up to multimegabar pressures", Phys. Rev. B 65, 064504 (2002).

**Acknowledgements**

The present work was supported by the National Key R&D Program of China under grant no. 2018YFA0305700, 2017YFA0302900 and 2015CB921300; the NSFC under grant no. 11474344. The work from HPSTAR was mainly supported by National Natural Science Foundation of China (Grant no. U1530402). Jinlong Zhu was supported by the National Thousand-Young-Talents Program.

**Additional information**

Supplementary Information accompanies this paper.

**Note:** The authors declare no competing financial interests.

Table I. Crystallographic data for $Ba_3TiTe_5$.

| Space group: $P6_3/mcm$ (193) - hexagonal | | | | | |
|---|---|---|---|---|---|
| $a = b = 10.1529(3)$ (Å), $c = 6.7217(1)$ (Å) | | | | | |
| $V = 600.05(9)$ (Å$^3$), $Z=2$ | | | | | |
| $\chi^2 = 2.0$, wRp = 4.4%, Rp = 3.1% | | | | | |
| Atom | Wyck. | x | y | z | Uiso |
| Ba | 6g | 0.61657(7) | 0.000000 | 0.250000 | 0.01650 |
| Ti | 2b | 0.000000 | 0.000000 | 0.000000 | 0.02351 |
| Te(1) | 6g | 0.23317(0) | 0.000000 | 0.250000 | 0.01202 |
| Te(2) | 4d | 0.333330 | 0.666667 | 0.000000 | 0.01762 |

Fig. 1. (a) The X-Ray diffraction pattern and the refinement of $Ba_3TiTe_5$. (b) and (c) are the schematic plots of the crystal structure with a view of projection along the *c* axis and perpendicular to the *c* axis, respectively. The atoms in the red dashed circles are Te chains on the Te(2) site. (d-f) The calculated band structure, PDOS and Fermi surfaces for $Ba_3TiTe_5$ under ambient pressure, respectively. (g) The resistivity as a function of temperature at ambient pressure. The inset is the curve of $\ln(\rho)$ versus $1/T$.

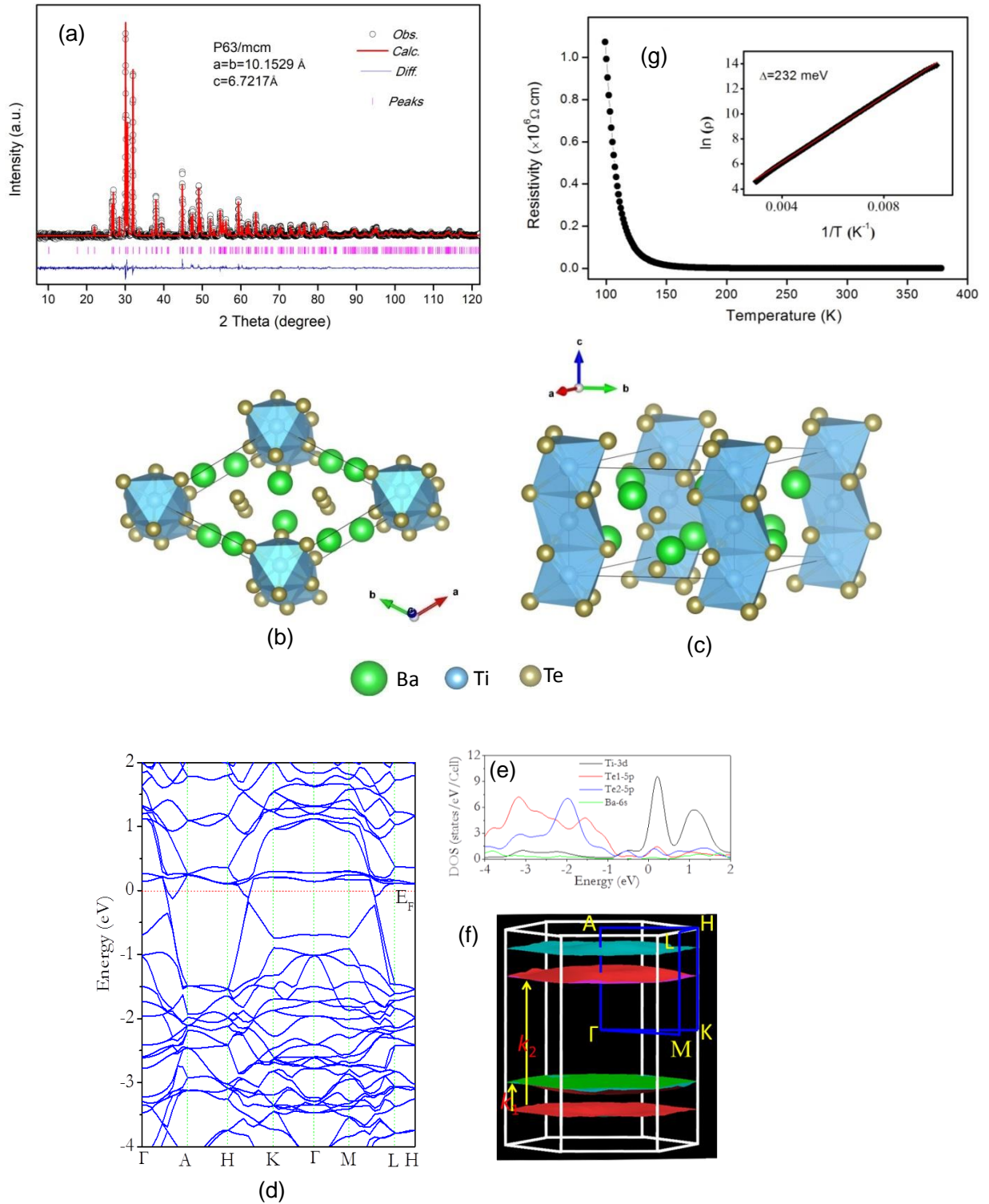

Fig. 2. (a) and (b) is the temperature dependence of resistance for $Ba_3TiTe_5$, measured under high pressure with the highest experimental pressure of 58.5 GPa. (c) The superconducting transition dependent on magnetic field at 17.3 GPa.

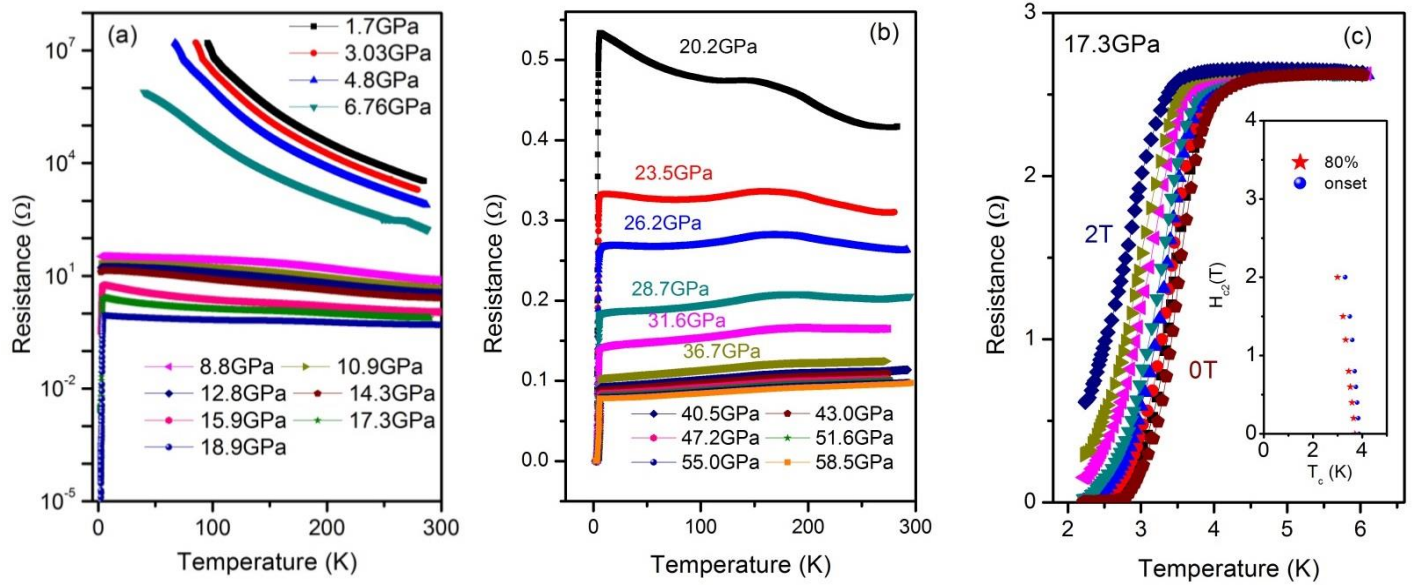

Fig. 3. The resistance curve for $Ba_3TiTe_5$ with an enlarged view of the pressure region from 18.9 GPa to 36.7 GPa shown in (a, b), and from 36.7 GPa to 58.8 GPa in (d, e). (c) The pressure dependence of the superconducting temperature $T_c$ and the MSC temperature $T_m$ in the range of 18.9 GPa to 36.7 GPa. (f) The pressure dependence of the superconducting temperature $T_c$ and the exponent $n$ in the range of 36.7 GPa to 58.8 GPa.

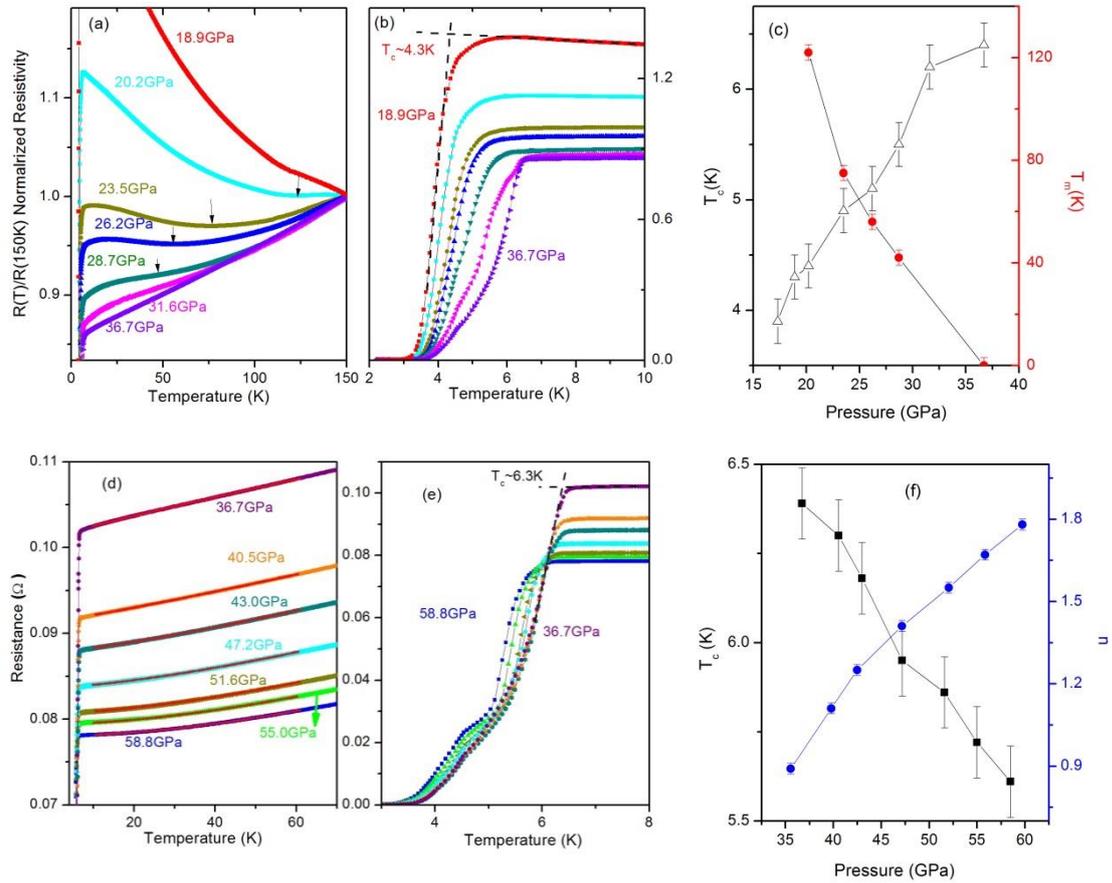

Fig. 4 (a-c) The calculated band structure, PDOS and Fermi surface for Ba$_3$TiTe$_5$ under 19.5 GPa. (d-f) The same as (a-c) under 42.2 GPa.

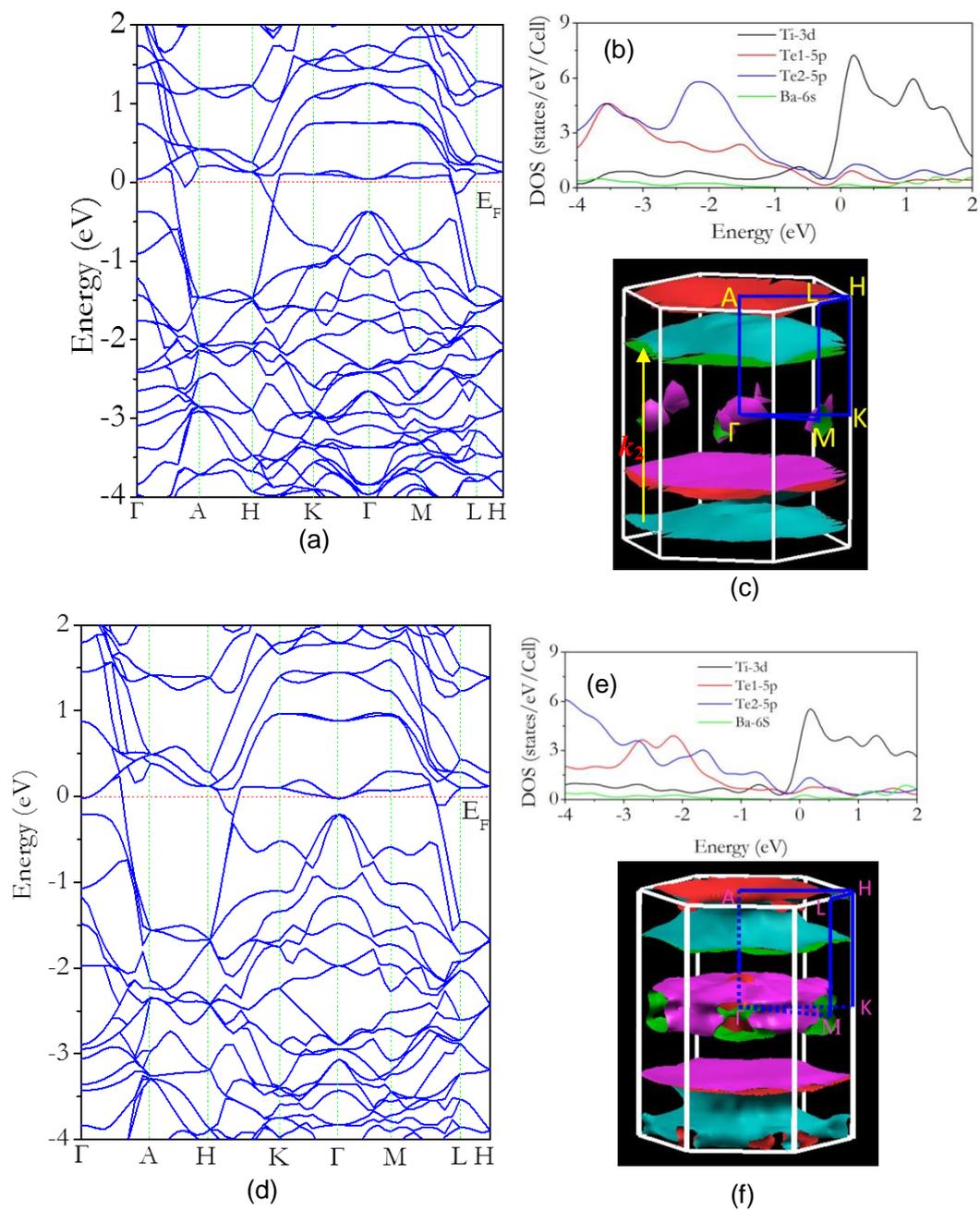

Fig. 5 The temperature-pressure phase diagram of $Ba_3TiTe_5$. The red circle denotes the superconducting temperature $T_c$, and the purple star is the MSC temperature $T_m$. In the metallic region, the color shading represents the value of exponent $n = d(\ln(R-R_0))/d(\ln(T))$.

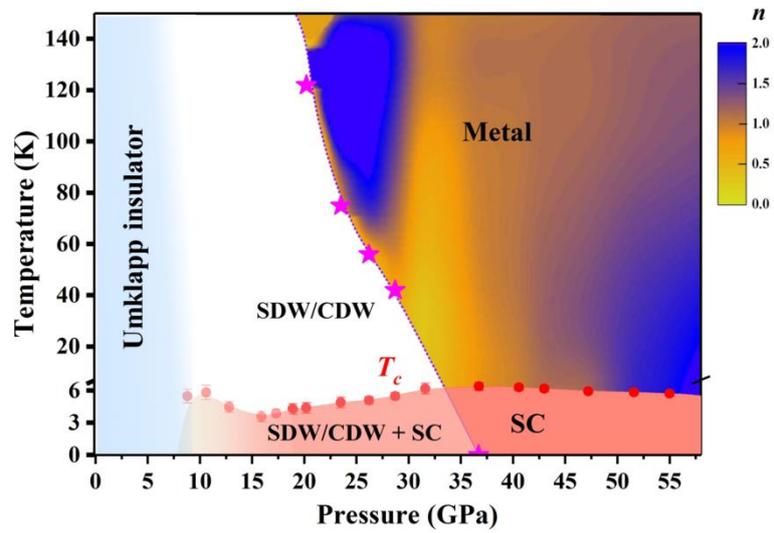